\documentclass[12pt]{article}
\usepackage{epsf} 

\def\be{\begin{equation}}
\def\ee{\end{equation}}
\def\bea{\begin{eqnarray}}
\def\eea{\end{eqnarray}} 
\def\nn{\nonumber \\}

\def\trho{{\tilde{\rho}}}
\def\ttheta{{\tilde{\theta}}}
\def\rc{{\rm cos}}
\def\rs{{\rm sin}}
\def\part{\partial}
\def\tfrac#1#2{{\textstyle{#1\over #2}}}

\def\cH{{\cal H}}

\def\makeatletter{\catcode`\@=11}
\makeatletter
\def\mathbox#1{\hbox{$\m@th#1$}}%
\def\math@ccstyles#1#2#3#4#5#6#7{{\leavevmode
     \setbox0\mathbox{#6#7}%
      \setbox2\mathbox{#4#5}%
      \dimen@ #3%
      \baselineskip\z@\lineskiplimit#1\lineskip\z@
      \vbox{\ialign{##\crcr
             \hfil \kern #2\box2 \hfil\crcr
             \noalign{\kern\dimen@}%
             \hfil\box0\hfil\crcr}}}}
\def\mathaccstyles{\math@ccstyles\maxdimen}
\def\maththroughstyles{\math@ccstyles{-\maxdimen}}
\def\unitmatrixDT%
 {\maththroughstyles{.45\ht0}\z@\displaystyle {\mathchar"006C}\displaystyle 1}
\begin{document}

\rightline{IFT-UAM/CSIC-99-20}
\rightline{hep-th/9905153}
\rightline{\today}
\vspace{1truecm}

\centerline{\Large{\bf Kaluza-Klein dipoles, brane/anti-brane pairs }}  
\vspace{.4cm}
\centerline{\Large{\bf and instabilities }}
\vspace{1truecm}

\centerline{{\bf Bert Janssen}\footnote{E-mail address: 
                                  {\tt bert.janssen@uam.es}}
{\bf and}
{\bf Sudipta Mukherji}\footnote{E-mail address: 
                                  {\tt sudipta.mukherji@uam.es}}}

\vspace{.4truecm}
\centerline{{\it Instituto de F{\'\i}sica Te{\'o}rica, C-XVI,}}
\centerline{{\it Universidad Aut{\'o}noma de Madrid}}
\centerline{{\it E-28049 Madrid, Spain}}
\vspace{2truecm}

\centerline{ABSTRACT}
\vspace{.5truecm}

\noindent
We review the method of constructing dipole and string loop solutions from the
higher-dimensional (Euclidean) Kerr black hole. We analyse the results in 
various dimensions, finding solutions earlier given in the literature. Then we
construct a new heterotic dipole with non-trivial dilaton and gauge 
fields. This can, in turn, be describes as a brane/anti-brane pair which 
interpolates between the KK-dipole and the $H$-dipole.
Finally an argument is presented on the tachyonic 
instability by analysing the string fluctuations on the dipole background.

\newpage
\noindent
{\large{\bf 1. Introduction}}

\vspace{.3cm}
Brane/anti-brane pairs have recently played very important
role in our understanding of stable non-BPS states in
string theory \cite{senrev}. As a classical vacuum, brane/anti-brane
pair leads to unstable configuration. First of all, they 
attract each other as they are oppositely charged. Moreover, 
when the separation between them is of the order of the
string length, there appears tachyon in the configuration
\cite{banks}. However, it has been argued in the 
recent past that, in certain
cases, this tachyon leads the system to a stable fixed point in the
renormalisation group sense \cite{sen3}. 

Since the classical force between brane/anti-brane system 
never vanishes, search for a configuration describing the pair 
as a classical solution looses its meaning. However,
if we introduce a non-zero background electromagnetic field
which triggers repulsion between brane and the anti-brane in such a
way that it cancels the attractive force between them, then one 
would expect to find a classical configuration describing the pair
in equilibrium\footnote{It may not be a stable equilibrium though.
Instabilities due to the tachyon is expected to be absent
if the distance between the pair is large enough.}. In \cite{sen},
such a solution was constructed in string theory and  various
instabilities were analysed. It has been argued that the Kaluza-Klein
(KK) dipole of 5-dimensional gravity \cite{gross} has the
property of describing brane/anti-brane pair when we add required 
flat directions\footnote{On the other hand, some of the 
dipole like solutions with magnetic flux tubes have directly been
constructed in \cite{davidson, sm}. However, most of
them do not have KK interpretation.}. 
An asymptotic magnetic field appears naturally
which, in turn, keeps the system in equilibrium. To our knowledge,
this is the only classical configuration known for the 
brane/anti-brane system till now.

Before we go into the detail of our analysis of brane/anti-brane 
pair let us first summarise what we do in this letter. 
In section 2.1, we review, in brief, the $D = 5$ dipole
following \cite{sen}. This dipole is constructed by starting with
Euclidean Kerr metric and adding a time direction. 
The dipole configuration is made out
of monopole, anti-monopole pair. While lifted up in eleven dimensions, it
describes a Dirichlet six and anti-six brane pair upon dimensional 
reduction over the Taub-NUT direction. In section
2.2, we ana\-lyse the situation in higher dimentions
\footnote{ This corrects a crucial interpretational error
that was made in an earlier version of this paper.}.
Following  \cite{gibbons} closely , we argue that 
similar construction in $D = 6$ gives
an oriented string loop. The opposite points on the loop behaves
as monopole or anti-monopole if we look close enough. We discuss
this in a very explicit manner. We further use these results in 
section 2.3 to excite gauge field along the world-sheet
direction of the string loop. Noticing the importance of 
brane/anti-brane configurations in understanding the physics of non-BPS
states, in section 3, we construct a very general five-dimensional
dipole which can be embedded in heterotic string theory. 
This solution is characterized  by an angular parameter. This parameter
allows the solution to interpolate between the usual KK dipole and 
the $H$-dipole.  We also, as
before, analyse the system critically to isolate each of the
constituents of the dipole system. 
In the last section, we try to understand if these brane/anti-brane 
configurations (in the presence of background electromagnetic field) can
be realised as stable string or superstring background at least when 
the distance between constituents is large. Unfortunately, we find that 
they likely lead to unstable string backgrounds. A string or superstring
propagating in these background contains tachyon in their
spectrum.

\vspace{1cm}
\noindent
{\large{\bf 2. Dipoles and Loops}}

\vspace{.3cm}
We start with the $(D-1)$-dimensional Kerr metric \cite{myers}:
\bea
dS^2_{(D-1)}&=& 
  -\ { r^2 + a^2 \rc^2\theta -2 M r^{6-D} \over {r^2 + a^2 \rc^2\theta}}
                               \ d\tau^2
    - {4Mr^{6-D}a \rs^2\theta\over{r^2 + a^2 \rc^2\theta}}\ d\tau d\phi
                                                   \label{Kerr}   \\  \nn  
&& + \ {\rs^2 \theta \over{r^2 + a^2 \rc^2\theta}}
        \Bigl[(r^2 + a^2)(r^2 + a^2 \rc^2\theta ) 
               + 2Mr^{6-D}a^2 \rs^2\theta\Bigr] \ d\phi^2 
                                                            \nn \nn
&& +\ (r^2 + a^2 \rc^2\theta)\Bigl[{dr^2 \over { r^2 + a^2-2 M r^{6-D}}} 
   + \  d\theta^2 \Bigr] 
   +\  r^2 \rc^2\theta \  d\Omega^{D-5} .  \nn \nonumber  
\eea
Note that this metric has horizon(s) when 
\be
r^2 + a^2 - 2 M r^{6-D} = 0. 
\label{Horizon}
\ee
For $D \le 6$, there is a critical value of $a$ beyond which the 
horizon does not exists. However, in case of $D > 6$, for any $a$ and
$M$, there exists one horizon. 

We then perform an Euclidean rotation
\be
\tau \rightarrow -iX, \hspace{1cm}
a \rightarrow ia,
\label{Euclrot}
\ee
and add a new time direction $t$. The solution thus obtained has the following 
form:
\bea
dS^2 &=& -dt^2 + (r^2 -  a^2 \rc^2\theta) (\frac{dr^2}{\Delta} + d\theta^2 )
         + \Delta \ \rs^2\theta \ \cH \ d\phi^2 
                                                             \nn
     && \hspace{1cm}+ {\cH}^{-1}\ (dX - A_\phi d\phi)^2 
       + r^2 \rc^2\theta \ d\Omega^{D-5},
\label{KKdipole}
\eea
where the gauge field $A_\phi$ and the functions $\cH$  are given 
by:
\bea
\cH=\frac{r^2 - a^2 \cos^2 \theta}{\Delta +a^2 \sin^2 \theta} ,
\hspace{1,5cm}
A_\phi= {2Mr^{6-D}a\ \rs^2\theta \over{\Delta + a^2 \rs^2\theta}}\ , 
\label{EuclKerr}
\eea
and
\be
\Delta = r^2 - a^2 - 2M r^{6-D},
\label{Delta} 
\ee

\noindent
Since the solutions (\ref{Kerr}) and (\ref{KKdipole}) are time independent,
the above construction is guaranteed to give a solution of the Einstein
equations and can be embedded in any theory containing gravity. 

\vspace{.5cm}

\noindent {\bf 2.1.~ Brane/anti-brane pair}

\vspace{.3cm}

In $D =5$, (\ref{KKdipole}) corresponds to monopole/anti-monopole
pair as has been discussed in \cite{sen}. This can be seen explicitly
by looking at the metric on $\Delta = 0$. At $\Delta = 0,$ and $\theta = 0$,
the above metric reduces to an anti-monopole configuration, and, on the
other hand, at $\Delta = 0$ and $ \theta = \pi$, we get a monopole 
configuration. Notice that, in order to avoid singularities, the
coordinates appearing in  (\ref{KKdipole}) have to lie within 
certain period. For $D = 5$, they are given by 
\bea
&&M + {\sqrt{M^2 + a^2}} \le r \le \infty, \ , \hspace{1cm}
0 \le X \le {4\pi M(M + \sqrt{M^2 + a^2})\over{\sqrt{M^2 + a^2}}}, \nn \nn
&& \hspace{3cm}
0 \le \phi - {aX\over {2M(M + {\sqrt{M^2 + a^2}})}} \le 2\pi.
\label{period}
\eea
The non-trivial periodicity of $\phi$ shows, in fact, the existence
of an external magnetic field, causing a repulsive magnetic force which 
cancels the attractive electric and gravitational forces and keeps 
the monopole and anti-monopole in equilibrium \cite{sen}. The distance
between the pair can be calculated easily by integrating the
metric along $\theta$. For large $a$, the distance turns out to be $2a$.
This associates a physical meaning to the parameter $a$.

The above configuration can be lifted up to $D=10$ or $D=11$ in a straight 
forward way by adding extra flat directions. Upon dimensional reduction
over $X$, the eleven dimensional metric reduces to 
 a Dirichlet six/anti-six brane pair in type
IIA string theory. Since under $T$-duality along $X$, the constituent
monopole goes to an $H$-monopole, we expect for $D = 5$, the $T$-dual
metric that follows from (\ref{KKdipole}) will describe an $H$-dipole.

\vspace{.5cm}
\noindent
{\bf 2.2.~ Oriented string loop}\footnote{We thank R. Empar\'an for pointing 
out a misinterpretation of this result in the previous version of this paper.} 
\vspace{.3cm}

One might wonder as to what happens if we generalize the dipole configuration 
of the sub-section 2.1 for higher dimensions without just adding flat 
directions. In other words, analysing Eqn.~(\ref{KKdipole}) for $D \ge 5$ in 
the same line as above. In that case, one gets a loop of KK-brane 
\cite{gibbons} {\it rather than} a monopole/anti-monopole. 
In this sub-section, we discuss the
configuration in some detail for $D = 6$. We will see that
indeed for $D = 6$, Eqn.~(\ref{KKdipole}) is nothing but a
KK-string loop lying along the transverse direction $d\chi$.
In the last sub-section, we will use this result to excite
gauge fields along the world-sheet direction of the loop.

As mentioned above, we start with Eqn.~(\ref{KKdipole}) for $D = 6$.
The metric in six dimensions takes the form:
\bea
dS^2 &=& -dt^2 + (r^2 -  a^2 \rc^2\theta) (\frac{dr^2}{\Delta} + d\theta^2 )
         + \Delta \ \rs^2\theta \ \cH \ d\phi^2 
                                                             \nn
     && \hspace{2cm}+ {\cH}^{-1}\ (dX - A_\phi d\phi)^2 
       + r^2 \rc^2\theta \ d\chi^2. 
\label{KK6}
\eea
where
\be
\Delta = r^2 -a^2 -2M \ , \hspace{1,5cm}
A_\phi = {2Ma \rs^2\theta\over {\Delta + a^2 \rs^2\theta}} \ .
\ee
To avoid the conical singularity at $r = \sqrt{a^2 + 2 M}$, 
one must have
\be
0 \le X \le {4\pi M\over{\sqrt{a^2 + 2 M}}} \ , \hspace{1cm}
0 \le \phi - {aX\over {2M}} \le 2\pi \ .
\label{KK6per}
\ee
Notice that in the limit of large $a$, the radius in the
$X$ direction depends on the ratio ${M\over a}$. This is unlike
the dipoles in $D = 5$, where the radius along $X$ at large $a$
is {\it independent} of $a$ \cite{sen}. 
Due to this twisted boundary condition (\ref{KK6per}) on $\phi$, there is 
an asymptotic magnetic field \cite{gibbons}. If we use $\psi =
\phi - aX/2M$ as independent coordinate, the magnetic field
is given by  $B={a}/{2M}$. \footnote{A different choice of coordinate will 
however change the magnetic field, see \cite{gibbons}.}

To have further insights of the configuration, we will now analyse
the metric at 
$\theta = 0, \pi$ when $\Delta$ vanishes. From the structure it is clear
that the zero of $\Delta$ occurs at $r = r_0 = \sqrt{a^2 + 2M}$.
To analyse the metric near $r = r_0, ~\theta = 0$, we define 
coordinates\footnote{This coordinate system turns out to be
the analogous one that was used in \cite{sen} for the five-dimensional dipole.}
\bea
&& {\sqrt{a^2 + 2M}}\ \rs^2 \theta = \tilde\rho\ (1 - \rc \tilde\theta),\\ \nn
&& 2(r - r_0) = \tilde\rho \ (1 + \rc \tilde\theta),
\label{NewCor}
\eea
and look at the limit 
\be
a \rightarrow \infty, \hspace{1cm}
M \rightarrow 0,   \hspace{1cm}
\theta \rightarrow 0, \hspace{1cm}
r \rightarrow  r_0 
\label{limit}
\ee
with $a \ \rs^2 \theta$ fixed and $r - r_0$ finite. In this limit, the metric
takes the following form
\be
dS^2 = - dt^2 + d \tilde \chi^2   
       + H^{-1} \Bigl( dX - A_\phi d\phi \Bigl)^2
       + H  \Bigl( d\tilde\rho^2    +    \tilde\rho^2 d\tilde\theta^2 
                    + \tilde\rho^2 \rs^2\tilde\theta d\phi^2\Bigr )
\label{KK}
\ee
with
\be
H = 1 + {\tilde M\over{\tilde\rho}} \hspace{1cm}
 A_\phi = \tilde M (1- \cos \tilde\theta) \hspace{1cm}
\tilde M = {M\over{\sqrt{a^2 + 2M}}}
\label{KKcomp}
\ee
In (\ref{KK}), 
$\tilde \chi = \sqrt{a^2 + 2M}\chi$ with period 
$0 \le \tilde\chi \le 2\pi\sqrt{a^2 + 2M}$. Thus we see that
near $r = r_0, \ \theta = 0$, there is a six-dimensional KK anti-monopole 
structure, a string-like object with world-sheet coordinates $(t, \tilde\chi)$.
Similarly, one can analyse at $r = r_0,\ \theta = \pi$. This gives a 
metric of a monopole configuration which has the 
same metric as (\ref{KK}) with a change of sign
in the gauge field. 

\begin{figure}
\begin{center}
\leavevmode
\epsfxsize=8cm
\epsffile{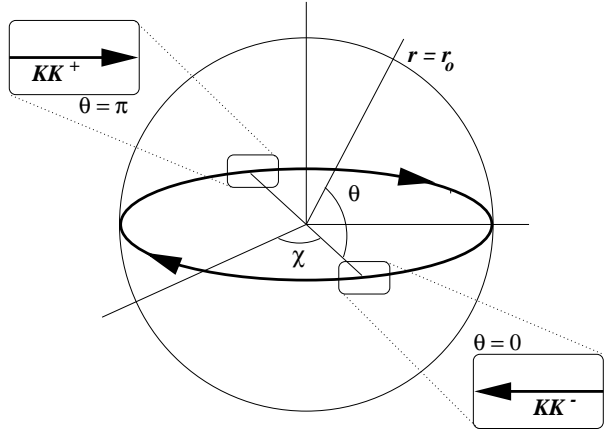}
\caption{\it A stringlike monopole wound around the $\chi$-direction. In the 
limit $r\rightarrow r_0$ and $\theta \rightarrow 0 \ (\pi)$ the solution looks like an anti-monopole (monopole). }
\label{loop}
\end{center}  
\end{figure}

However, we must notice that the $\chi$-direction is an isometry direction with
period $2\pi r_0$. Hence the limit (\ref{limit}) is true for every value of 
$\chi$ and what we are seeing is actually an oriented string loop along 
$\chi$ (see Figure \ref{loop}).  The (anti)-monopole structure (\ref{KK}) in 
the limit (\ref{limit}) is just a limit artifact,  the different signs of the 
gauge fields at $\theta = 0$ and $\theta= \pi$ is due to the opposite 
orientation of the loop.


\vspace{.5cm}

\noindent
{\bf 2.3.~ Exciting  gauge field on the loop}

\vspace{.3cm}

In general, a $D$-dimensional Kerr black hole is characterized by 
$[\tfrac{D-1}{2}]$ angular momentum parameters \cite{myers}, where the square 
brackets indicate the integer part. In order to obtain a string loop  
configuration with more then one non-trivial gauge field, we need to start 
with the general $D$-dimensional Kerr metric. We illustrate 
this by starting with the Kerr metric in five dimensions with two non-zero 
angular momenta proportional to $a_1$ and $a_2$ \cite{myers}. We then 
follow the same procedure as described in  (\ref{Euclrot}). The resulting
metric has the form:
\bea
dS^2 &=& -dt^2 + (r^2 - a_1^2 \rc^2\theta - a_2^2 \rs^2\theta )
\Bigl [ {dr^2\over{\tilde\Delta + a_1^2a_2^2 r^{-2}}} + d\theta^2 \Bigr ]
                                             \nn \nn
&+&\frac{\rs^2\theta
         [(r^2 - a_1^2)(\tilde\Delta + a_1^2\rs^2\theta + a_2^2
\rc^2\theta)
                 - 2 M a_1^2\rs^2\theta]}
        {\tilde\Delta + a_1^2 \rs^2\theta + a_2^2 \rc^2\theta} \ d\phi^2
                                     \nn \nn
&+&\frac{\rc^2\theta
         [(r^2 - a_2^2)(\tilde\Delta + a_1^2\rs^2\theta + a_2^2
\rc^2\theta)
                - 2 M a_2^2\rc^2\theta]}
        {\tilde\Delta + a_1^2 \rs^2\theta + a_2^2 \rc^2\theta} \ d\chi^2 
                                      \nn \nn
&-& \frac{4a_1a_2 M \rs^2\theta \rc^2\theta}
         {\tilde\Delta + a_1^2\rs^2\theta + a_2^2 \rc^2\theta} \  d\phi
d\chi
                                     \nn  \nn
&+&\frac{\tilde\Delta + a_1^2\rs^2\theta + a_2^2 \rc^2\theta}
        {r^2 - a_1^2 \rc^2\theta - a_2^2 \rs^2\theta} \
           \Bigl [ dX - A_\chi d\chi - A_\phi d\phi \Bigr ]^2
\label{GeneuKerr}
\eea
with $\tilde\Delta = r^2 - a_1^2 - a_2^2 - 2M$ and
\be
A_\chi = \frac{2Ma_2 \rc^2\theta}
              {\tilde\Delta + a_1^2\rs^2\theta + a_2^2 \rc^2\theta} \ ,
\hspace{1cm}
A_\phi = \frac{2Ma_1 \rs^2\theta}
              {\tilde\Delta + a_1^2\rs^2\theta + a_2^2 \rc^2\theta} \ .
\label{GenGauge}
\ee

The general analysis of this metric is very complicated and it is not clear 
what the interpretation of the above solution is. 
We will restrict ourselves to the case $a_1 \gg a_2$. In this 
case, the metric structure simplifies and can be analysed in
a straight forward manner. However, in doing so, we will miss
out important physics which appear at order ${\cal O}(a^2)$ or
higher.

In the limit $a_1 \gg a_2$, (\ref{GeneuKerr}) reduces to the
following simple form:
\bea
dS^2 &=& -dt^2 + (r^2 - a_1^2 \rc^2\theta)\Bigl [ {dr^2\over \Delta} +
       d\theta^2\Bigr ]
                        \nn \nn
     &-& {4a_1a_2 M\rc^2\theta \rs^2\theta\over
      {\Delta + a_1^2 \rs^2\theta}}             \ d\phi d\chi +
      {\Delta\rs^2\theta (r^2 - a_1^2\rc^2\theta )
       \over {\Delta + a_1^2 \rs^2\theta}}      \ d\phi^2
                                     \nn \nn
     &+& r^2\rc^2\theta d\chi^2 +
     {\Delta + a_1^2 \rs^2\theta\over{(r^2 - a_1^2 \rc^2\theta)}}
      (dX - A_\chi d\chi - A_\phi d\phi )^2,
\label{SimKerr}
\eea
with $\Delta = r^2 - a_1^2 - 2M$ and 
\be
A_\chi = {2Ma_2 \rc^2\theta \over{\Delta + a_1^2\rs^2\theta }}, 
\hspace{2cm}
A_\phi = {2Ma_1 \rs^2\theta \over{\Delta +  a_2^2
\rc^2\theta}}.
\label{SimGauge}
\ee
In the limit  (\ref{limit}), the metric reduces to 
\bea
dS^2 &=& -dt^2 + d\tilde\chi^2 
 +  H(d\tilde\rho^2 + \tilde\rho^2 d\tilde\theta^2 
                    + \tilde\rho^2\rs^2\tilde\theta d\phi^2) 
                                  \\ \nn
&& \hspace{1cm} 
       - 2 {\tilde a}_2 {\tilde M}(1 - \rc\tilde\theta)\ d\tilde\chi d\phi 
     \ + \ H^{-1}(dX - A_\phi d\phi - A_{\tilde\chi} d\tilde\chi)^2 \nonumber
\eea
with the expression for the gauge fields given by: 
\be
A_\phi = \tilde M (1 - \rc\tilde\theta ), \hspace{2cm}
A_{\tilde\chi} = {{\tilde M} {\tilde a}_2\over {\tilde\rho}}.
\label{NearHor}
\ee
Here the parameters $r_0, \ \tilde M$ have the same form as before
with $a$ replaced by $a_1$ and ${\tilde a}_2 = a_2/{\sqrt{a_1^2 + 2M}}$.
Comparing with (\ref{KK}) and (\ref{KKcomp}), we see that the
above metric describes a string loop with gauge field $A_\chi$ excited
on the world-sheet direction $\chi$.

In exactly similar manner, we can study the limit $a_2 \gg a_1$.
In this case, the poles change the position, now located 
at $\theta=\pi/2$ and , $3\pi/2$ and the gauge fields interchange their
roles. 

\vspace{1cm}

\noindent
{\large\bf{3. Heterotic Dipole}}

\vspace{.3cm}

A more general dipole solution can be found, starting from the general rotating
charged black hole solution of $D=4$ heterotic string theory. Such a solution 
with one $U(1)$ gauge field $V_\mu$ (all other $U(1)$ vectors, moduli and 
vectors coming from compactification set equal to zero) was given in 
\cite{sen2}. The metric is characterized by three parameters, namely $M, a$ 
and $\alpha$, where $\alpha$ parametrizes an $O(1,1)$ $T$-duality rotation 
between the Kerr-solution and the heterotic charged rotating black hole. 
After an Euclidean rotation
\be
\tau \rightarrow -iX, \hspace{1cm}
a \rightarrow ia,  \hspace{1cm}
\alpha  \rightarrow i\alpha,
\label{Euclrot2}
\ee
and adding a new time direction $t$, we get the following solution of
$D=5$
heterotic string theory (in the string frame, following the notation of 
\cite{BHO}):
\bea
dS^2 &=& -dt^2 + (r^2 -  a^2 \rc^2\theta) (\frac{dr^2}{\Delta} + d\theta^2 )
         + \Delta \ \rs^2\theta \ \cH_{(\alpha = 0)} \ d\phi^2 
                                                             \nn
     && \hspace{4cm}+ {\cH}_{(\alpha)}^{-1}\ 
                   \Bigl[dX - A^{(\alpha)}_\phi d\phi \Bigr]^2 
                                 ,\label{KKalpha} \\ \nn
e^{-2\Phi_{(\alpha)}}
          &=&\frac{r^2 - a^2\cos^2\theta -2Mr\sin^2\tfrac{\alpha}{2}}
                  {r^2 - a^2\cos^2\theta} \ ,
\ \ \ 
B_{X\phi}^{(\alpha)} = \frac{2Mra \sin^2\theta \sin^2\tfrac{\alpha}{2}}
                   {r^2 - a^2\cos^2\theta -2Mr\sin^2\tfrac{\alpha}{2}}\nn\nn
V_X^{(\alpha)} &=& \frac{-2Mr  \sin\alpha}
                   {r^2 - a^2\cos^2\theta -2Mr\sin^2\tfrac{\alpha}{2}}\ ,
\ \ \
V_\phi^{(\alpha)} = \frac{-2Mra \sin^2\theta  \sin\alpha}
                   {r^2 - a^2\cos^2\theta -2Mr\sin^2\tfrac{\alpha}{2}} 
\nonumber
\eea
where the functions $\cH_{(\alpha)}$ and $A^{(\alpha)}_\phi$ are given by
\bea
\cH_{(\alpha)}&=& \frac{(r^2 - a^2\cos^2\theta-2Mr\sin^2\tfrac{\alpha}{2})^2}
                     {(\Delta + a^2\sin^2\theta)(r^2 - a^2\cos^2\theta)}\ , 
                                             \nn \nn
A^{(\alpha)}_\phi &=&\frac{2Mra\sin^2\theta\cos^2\tfrac{\alpha}{2}}
                          {\Delta + a^2\sin^2\theta}\ .
\eea
Note that for $\alpha = 0$ we recover the five-dimensional dipole
(\ref{KKdipole}), and for $\alpha = \pi$, the solution takes the form of a 
$S0/{\rm anti-}S0$-brane ($H$-dipole) which is $T$-dual to (\ref{KKdipole}). 
The dipole nature of the interpolating solution can be easily seen from the 
asymptotic behaviour of the vector fields $A^{(\alpha)}_\phi$ and 
$B_{X\phi}^{(\alpha)}$. The total dipole moment of the solution is $2Ma$, to 
which the above gauge fields contribute with a factor $\cos^2\tfrac{\alpha}{2}$
and $\sin^2\tfrac{\alpha}{2}$ respectively .
Furthermore, to avoid singularities associated with the above metric, one
must have

\bea
&&M + {\sqrt{M^2 + a^2}}\  \le r \le \infty \ ,
                                       \nn\nn
&&0\ \le X \le {2\pi M (1+ \cos \alpha )(M + {\sqrt{M^2 + a^2}} )
\over {  {\sqrt{M^2 + a^2}}}} \ ,
                             \nn\nn
&&0\ \le \phi - {aX\over{M (1+ \cos\alpha )(M + {\sqrt{M^2 + a^2}} )}}
\le 2\pi \ .
\eea
Note that for $\alpha = 0$ we find the periodicity of the previous case 
(\ref{period}), but howeverwe do not understand the limit $\alpha \rightarrow 
\pi$, where the period of $X$ goes to zero.   
As before, due to the non-trivial periodicity of $\phi$, we see
an appearance of background magnetic field.
In the near pole limit (\ref{limit}), the solution (\ref{KKalpha}) is of the 
form
\bea
dS^2 &=& -dt^2 
          + \frac{H_{(0)}}{H_{(\alpha)}^2} 
                      \Bigl[dX + A_\phi^{(\alpha)}d\phi \Bigr]^2
          + H_{(0)} \Bigl[ d\trho^2  + \trho^2 d\ttheta^2 
                                  +\trho^2 \sin^2\ttheta d\phi^2\Bigr]\ , 
                                                \nn\nn
e^{-2\Phi_{(\alpha)}} &=& \frac{H_{(0)}}{H_{(\alpha)}}\ ,    \hspace{3cm}
B_{X\phi}^{(\alpha)} = 
       - H_{(\alpha)}^{-1} ( A_\phi^{(\alpha)} -  A_\phi^{(0)} ) , \\ \nn
V_\phi^{(\alpha)} &=&  H_{(\alpha)}^{-1}  A_\phi^{(0)} \sin \alpha  \ , 
                           \hspace{1.5cm}
V_X^{(\alpha)} = - H_{(\alpha)}^{-1}\ \frac{M}{\trho}\ \sin \alpha\ , \nonumber
\eea
where
\be
H_{(\alpha)}= 1 + \frac{M}{\trho} \cos^2\tfrac{\alpha}{2} \ ,\hspace{1cm}
A_\phi^{(\alpha)} = M(1-\cos\ttheta)\cos^2\tfrac{\alpha}{2} \ .
\ee

The geodesic distance among the constituents of the dipole can be
calculated from the metric by integrating along $\theta$. For 
large $a$, the distance turns out to be $2a$ in the string-frame.

\vspace{1cm}
\noindent
{\large\bf{4. Tachyonic Instability}}

\vspace{.3cm}
All the solutions that are discused in (\ref{KKdipole})
can be embedded obviously in any string or superstring
theories since they are the solutions of Einstein gra\-vity.
As a consequence, they can be considered as various 
string backgrounds. A natural question  thus  is to ask if
these backgrounds are stable. First thing in this direction
will be to check if string/superstring propagating in these 
backgrounds contain tachyons in their fluctuation spectrum. 
The aim of this section is to carry out an analysis of this 
issue. In general, string fluctuations 
in a non-trivial background is hard to analyse. This is
because one needs to have a description of the background in
terms of world-sheet conformal field theory (CFT). 
The dipole backgrounds
that we have discussed earlier have very complicated field
configurations. However, at very large radial distance the 
the structure simplifies. Fortunately for us, in this regime,
the two dimensional CFT is known and 
has been analysed in the literature \cite{tseytlin}.
We will thus make use of his results.
 
As an illustrative example, we will work with the
solution (\ref{KKdipole}) for $D = 5$. The solution can 
be read off from (\ref{KKdipole}). Various parameters 
are 
\bea
\Delta &=& r^2 - 2Mr - a^2,~~~r_0 = M + \sqrt{M^2 + a^2}.
\eea
Here as before $r_0$ corresponds to the zero of $\Delta$.
For our present purpose, we will also need the periods of
$X$ and $\phi$. Their ranges are
\bea
0 \le X \le {4\pi M(M + \sqrt{M^2 + a^2})\over{\sqrt{M^2 + a^2}}},~~~
0 \le \phi \le {2\pi a\over {\sqrt{M^2 + a^2}}}.
\label{Period}
\eea
In order to proceed, we first notice that for $r \rightarrow \infty$,
the metric reduces to 
\be
dS^2 = - dt^2 + d\rho^2 + dz^2 + \rho^2 d\phi^2 + dX^2.
\ee
Here we have defined $\rho = r \rs\theta, z = r \rc\theta$.
Thus the metric is flat. However, due to nontrivial periods
of various coordinates (\ref{Period}), there is an asymptotic 
magnetic field $B$ given by
\be
B = {a\over {2M (M + {\sqrt{M^2 + a^2}})}}.
\ee
Upon reduction over $X$, we would thus get four dimensional 
Melvin solution (see \cite{dowker} for detail).

Fortunately, Melvin background is  one of the very few 
backgrounds where string world-sheet has a description
in terms of CFT \cite{tseytlin}\footnote{Of course, 
to have the right central charge, one needs to add to this
background an internal CFT or SCFT. Since this internal CFT
will be completely decoupled from space-time physics, we will 
not pay attention to this sector.}. 
If we denote the radius of $X$ by $R$ (which follows from
(\ref{Period})), then for integer $1\over{BR}$, one 
has an orbifold CFT. The non-trivial part of the
CFT is a $Z_N$ orbifold of $2$-dimensional plane times 
a circle, where $N$ is related to the magnetic field of the 
Melvin solution. However, it is known that this CFT
description for string or superstring, contains tachyon
in the spectrum and the mass formula is given by
$\alpha^\prime m^2 = -4 + {4\over N}$ \cite{lowe}. From this observation,
we thus conclude that the dipole solution for $D = 5$
at large radial distance leeds to unstable string or 
superstring background due to the presence of tachyon.


\vspace{1cm}
\noindent
{\Large{\bf Acknowledgments}} 

\vspace{.2cm}
\noindent
We have grately benefited from continuous discussion with C\'esar
G\'omez on brane/anti-brane systems. We also like to thank Tom\'as Ort{\'\i}n 
for useful discussions and especially Roberto Empar\'an for pointing out a 
crucial mistake in an earlier version of this paper.
The work of B.J.~has been supported by the TMR program FMRX-CT96-0012 and the 
work of S.M.~by the Ministerio de Educaci\'on y Cultura of Spain and the 
grant CICYT-AEN 97-1678.


\end{document}